\documentclass[%
 aip,
 amsmath,amssymb,
 reprint,%
]{revtex4-1}

\usepackage{graphicx}
\usepackage{dcolumn}
\usepackage{bm}

\usepackage[utf8]{inputenc}
\usepackage[T1]{fontenc}
\usepackage{mathptmx}
\usepackage{etoolbox}

\usepackage{hyperref}

\makeatletter
\def\@email#1#2{%
 \endgroup
 \patchcmd{\titleblock@produce}
  {\frontmatter@RRAPformat}
  {\frontmatter@RRAPformat{\produce@RRAP{*#1\href{mailto:#2}{#2}}}\frontmatter@RRAPformat}
  {}{}
}%
\makeatother
\begin{document}

\preprint{AIP/123-QED}

\title[Open Ferromagnetic Resonance Broadband Spectrometer]{OpenFMR: A low-cost open-source broadband ferromagnetic resonance spectrometer}
\author{Markus Meinert}
 \email{markus.meinert@tu-darmstadt.de}
\author{Tiago de Oliveira Schneider}
\author{Shalini Sharma}%
\author{Amir Khan}%
\affiliation{ 
New Materials Electronics Group, Department of Electrical Engineering and Information Technology, Technical University of Darmstadt, Merckstr. 25, 64283 Darmstadt, Germany
}%


\date{\today}

\begin{abstract}
We describe a broadband ferromagnetic resonance spectrometer for scientific and educational applications with a frequency range up to 30\,GHz. It is built with low-cost components available off-the-shelf and utilizes 3D printed parts for sample holders and support structures, and requires little assembly. A PCB design for the grounded coplanar waveguide (GCPW) is presented and analysed. We further include a software suite for command-line or script driven data acqusition, a graphical user interface, and a graphical data analysis program. The capabilities of the system design are demonstrated with measurements on ferromagnetic thin films with a thickness of 1\,nm. All designs and scripts are published under the GNU GPL v3.0 license.

\end{abstract}

\maketitle

\section{\label{sec:intro}Introduction}
The ferromagnetic resonance (FMR)\cite{GRIFFITHS1946, Kittel1948} is a key property of magnetization dynamics and is described by the Landau-Lifshitz-Gilbert (LLG) equation:
\begin{equation}\label{eq:LLG}
\frac{d\bm{m}}{dt} = - \gamma \left(\bm{m} \times \bm{B}_\mathrm{eff}\right) + \alpha\left( \bm{m} \times \frac{d\bm{m}}{dt} \right).
\end{equation}
Here, $\gamma$ is the gyromagnetic ratio and $\alpha$ is the Gilbert damping parameter with typically $\alpha \approx 0.001 \dots 0.1$. The equation describes the motion of the magnetization unit vector $\bm{m} = \bm{M} / M_\mathrm{s}$ (with the saturation magnetization $M_\mathrm{s}$) in response to a magnetic field (external or internal) and the magnetization damping due to energy dissipation. The LLG predicts a resonance condition where at a given effective magnetic field $\bm{B}_\mathrm{eff}$ (we use the convention $B=\mu_0 H$ and refer to the external flux density) a precession of the magnetization can be excited with a resonant magnetic field of frequency $f_0$. The FMR allows for a direct measurement of the Gilbert damping constant $\alpha$ via inspection of the resonance linewidths\cite{Patton1968, Kalarickal2006}, and gives indirect access to the Land\'e $g$-factor and the magnetization $M_\mathrm{s}$ by fitting the resonance position to the Kittel equation\cite{Kittel1948}. For thin films, the technique can be applied to both in-plane and out-of-plane magnetization, where we get the following two special cases for the resonance condition:

\begin{align}
f_\mathrm{res} &= \gamma\,' \sqrt{B \left(B+\mu_0 M_\mathrm{eff}\right)} \qquad {\text{(in-plane),}} \label{eq:Kittel_ip}\\
f_\mathrm{res} &= \gamma\,' \left( B - \mu_0 M_\mathrm{eff}\right) \qquad {\text{(out-of-plane).}} \label{eq:Kittel_oop}
\end{align}
The effective magnetization includes the perpendicular anisotropy field, $M_\mathrm{eff} = M_\mathrm{s} - \frac{2K_\perp}{\mu_0 M_\mathrm{s}}$. More complicated expressions are found when in-plane anisotropies (e.g. uniaxial, cubic, ...) are taken into account. Because of the reduced gyromagnetic ratio $\gamma\,' = \gamma / 2\pi = ge / 4\pi m_e \approx 29.4$\,GHz/T (with $g \approx 2.1$), the resonance frequencies are typically found between 1\,GHz and 60\,GHz.

Traditionally, there are two approaches for measuring the ferromagnetic resonance via microwave absorption:\cite{Lo2013, Montoya2014, Maksymov2015, Mewes2021} a) by keeping the frequency fixed and sweeping the magnetic field; and b) by keeping the magnetic field fixed, and sweeping the frequency. The fixed-frequency measurements can be implemented with resonant cavities (often at 9.4\,GHz), or with broadband microstrips or coplanar waveguides with the film placed directly on top of the line. The same microstrips and coplanar waveguides can be used for frequency-swept measurements, but come with a huge disadvantage: due to impedance mismatch and radiative losses, the transmission through the waveguide can have sharp dips and therefore requires careful comparison between measurements with and without the sample. In contrast, the field-swept measurements are easy to implement and insensitive to the specific transmission properties of the waveguides.

In the simplest case, the line shape of the absorbed power $\Delta P(B)$ in the field-swept experiment can be expressed as a Lorentzian line shape
\begin{equation}\label{eq:Lorentz}
\Delta P_\mathrm{L}(B) = A \frac{\Gamma_B^2}{\left(B_\mathrm{res} - B\right)^2 + \Gamma_B^2 }.
\end{equation}
The linewidth parameter $\Gamma_B$ is the half width at half maximum (HWHM) of the Lorentzian. It is related to the Gilbert damping parameter $\alpha$ and the frequency as\cite{Patton1968, Kalarickal2006}
\begin{equation}\label{eq:width}
\Gamma_B(f) = \alpha \frac{f}{\gamma\,'} + \Delta B(0)
\end{equation}
with the inhomogeneous broadening $\Delta B(0)$, which arises from inhomogeneities in the sample and thereby smears out the resonance over the probed sample volume.

\begin{table*}
\caption{\label{tab:shoppinglist}The shopping list. Prices are given without VAT according to quotations from years 2022 through 2024.}
\begin{ruledtabular}
\begin{tabular}{llr}
Component&Details&Price\\ \hline
RF Signal Generator	& DSI SG30000PRO low-noise signal generator, 30GHz, 15.5\,dBm output power			&	\$ 5,200\\
Detection Diode		& Krytar 203BK unbiased Schottky diode, 2.92mm to BNC							&	\$ 600\\
RF Cables			& Thorlabs KMM24, 2.92mm													&	\$ 250\\
CPW					& co-planar waveguide with ground on 200\,$\mathrm{\mu}m$ Rogers 4003C with FR4 support	&	\$ 700 / 10pc\\
Connectors			& WithWave 2.92mm End Launch Connectors (Narrow Block), 40GHz rated						&	\$ 80 / pc\\\hline
Electromagnet			& Xiamen Dexing Magnet Tech DXWD-80, 80mm core, C-Frame					&	\$ 5000\\
Power Supply			& Caenels Fast-PS-1k5 50V 30A                                          				&	\$ 9500\\\hline
Modulation Coils		& 0.20 mm wire, 2 $\times$ 50 turns, Helmholtz configuration, 32 mm diameter, 3D printed supports				& \$ 20\\
Modulation Amplifier	& Kemo M032S 12W mono audio amplifier										& \$ 20\\\hline
Lock-In Amplifier		& Zurich Instruments MFLI													& \$ 7000\\
Gaussmeter			& Magnet-Physik FH 55 Tabletop Gaussmeter									& \$ 2500\\
\end{tabular}
\end{ruledtabular}
\end{table*}

\begin{figure}
\includegraphics[width=8.6cm]{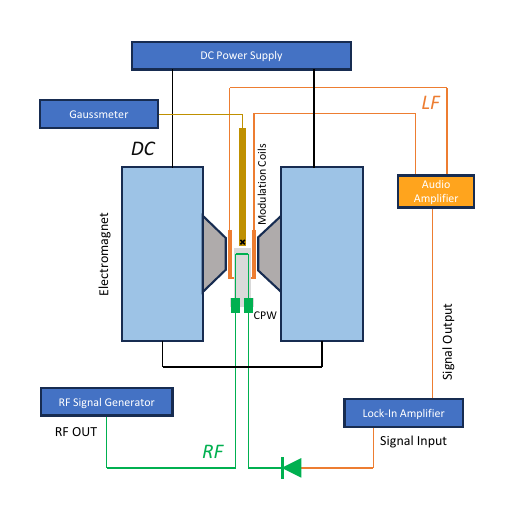}
\caption{\label{fig:setup}Schematic of our broadband field-sweeping FMR spectrometer showing its essential radio frequency, low frequency, and DC components. Dark blue components are controlled by our software from the computer.}
\end{figure}

While FMR measurements at a single frequency can give insight into the magnetization dynamics of a thin-film system and are very suitable for the study of angular dependences and magnetic anisotropy, broadband measurements give access to a more complete picture and are mandatory for unambiguous fitting of $g$, $M_\mathrm{eff}$, $\alpha$, and $\Delta B_0$. Broadband measurements are either done with a vector network analyzer (VNA), or with an RF signal generator and a diode for signal rectification and detection. The VNA performs fast, frequency-swept measurements and allows the collection of 2D datasets of the frequency- and field-dependent microwave absorption through the sample. However, these devices are relatively costly for a frequency range up to 30\,GHz or even more and cost easily around 35,000 USD. High frequencies up to 30\,GHz or more are desirable to improve the fit accuracy for both $M_\mathrm{eff}$ and $g$, especially in the in-plane configuration.\cite{Shaw2013, Zhang2017, Gonzalez2018} In contrast to a VNA, a simple RF signal generator and a diode are substantially cheaper. However, the absorption signal is relatively weak and needs to be improved via phase-sensitive detection with a lock-in amplifier, that is locked to the modulation frequency of a weak ac magnetic field at a few hundred Hertz, which is overlaid in parallel with the dc magnetic field being swept. Instead of the Lorentz line shape of Eq. \eqref{eq:Lorentz}, a field-derivative of the line shape is measured. This detection scheme requires that we add the cost for a lock-in amplifier to the total cost of the setup. Turn-key solutions that implement this detection scheme are commercially available, but are even more expensive than a VNA. More recently, hybrid concepts of field-modulated VNA detection have been developed to further enhance the signal-to-noise ratio of the VNA detection.\cite{Tamaru2018, Tamaru2021}

In the following, we will outline which components are necessary for the construction of a field-sweeping FMR spectrometer and we discuss our individual choices. The Supplemental Material contains a list of possible alternatives. Then, we briefly outline our software and design choices made here. Finally, we show examples of FMR measurements on ferromagnetic thin films to demonstrate the capabilities of our setup.

\section{\label{sec:hardware}Hardware selection}
In Figure \ref{fig:setup} we display a schematic of our setup. We point out that it is built with simplicity in mind and uses a minimal number of components. Table \ref{tab:shoppinglist} shows our shopping list for building the full FMR spectrometer. However, in many magnetism laboratories most of these components will already be available and the FMR spectrometer can be built by just adding the radio-frequency (RF) and low-frequency (LF) components to existing instrumentation.

The RF instrumentation comprises a low-cost, yet low-noise and high-power compact signal generator with an output up to 30\,GHz via a 2.92\,mm (K) connector from DS Instruments; this instrument is the cheapest device we could find on the market after an extensive search, yet it has very good specifications. The same manufacturer offers other similarly priced instruments with frequency ranges up to 40\,GHz that can be chosen if the focus of the measurements lies on a reliable determination of  $M_\mathrm{eff}$ and $g$ in the in-plane configuration.\cite{Shaw2013, Zhang2017, Gonzalez2018} We chose relatively low-cost cables, which are sufficient given that they can be short and the damping does not play a big role here. However, an upgrade to higher-quality cables with lower damping is recommended if longer cables are desired. For the RF rectification and detection we chose a packaged Schottky diode to simplify its handling.

The electromagnet must be able to sustain a homogeneous magnetic field of up to about 800\,mT for typical magnetic thin films in the in-plane configuration across a 20\,mm gap to accomodate the CPW. For the out-of-plane configuration with in-plane magnetized thin films, the magnetic field has to be much stronger, so we reduce the gap to 5\,mm to obtain fields of more than 2\,T. The field homogeneity in the in-plane configuration is crucial for an accurate measurement of the linewidth, as an inhomogeneous field leads to an artificial broadening of the linewidth with large samples. A magnet with a pole diameter of at least 75\,mm to 80\,mm and tapered poles is thus recommended. The magnet can be operated with a unipolar power supply. However, because of the magnet's remanence, it is helpful to use a bipolar power supply to obtain an accurate zero magnetic field for lower frequency measurements in samples with small damping. Our selected power supply is a bipolar power supply with current and voltage ratings adapted to our magnet, which is operated in a parallel coil configuration. By rewiring it to a serial coil configuration (or ordering it from the manufacturer like this), cheap programmable unipolar power supplies can be used instead. The modulation coils were wound by hand on a 3D printed support structure that is simply clamped between the magnet poles. We designed two different modulation coils, optimized for in-plane and out-of-plane operation, respectively, matching the 20\,mm and 5\,mm pole gaps. A cheap mono audio amplifier is used to drive the modulation coils with a sinusoidal output signal from the lock-in amplifier and is adjusted to provide a sinusoidal modulation of about 1\,mT RMS (2 mT at 5mm gap) with minimal distortion. The field measurement has to be very accurate to ensure correct interpretation of the FMR measurements. Thus, we chose a calibrated tabletop gaussmeter with a 0.4\% DC accuracy. 

The lock-in amplifier plays a crucial role for the signal extraction of the field-modulated resonance measurement. It generates the signal for the modulation coils and demodulates the rectified voltage. The Zurich Instruments MFLI device is an excellent choice for this task, as it provides many additional functions, differential outputs and inputs, auxiliary outputs and inputs, multiple digital instruments (such as a oscilloscope and FFT for modulation frequency selection or noise spectrum analysis), and excellent noise characteristics. However, even with a cheap do-it-yourself lock-in amplifier like the Open Lock-In Amplifier (OLIA),\cite{Harvie2023} high quality measurements can be obtained. To account for phase shifts, we perform automatic phase optimization via rotation of the signal in the complex plane to maximize the in-phase component after each field sweep.

For many components that we chose one can find alternatives. However, to the best of our knowledge, these are either the cheapest or best-value components available. The Supplemental material contains further details about our hardware and the 3D prints. In the next section, we briefly outline the design process of the coplanar waveguide and the software that we wrote for the measurement system.

\section{\label{sec:CPW}Grounded Coplanar Waveguide}

\begin{figure}
\includegraphics[width=8.6cm]{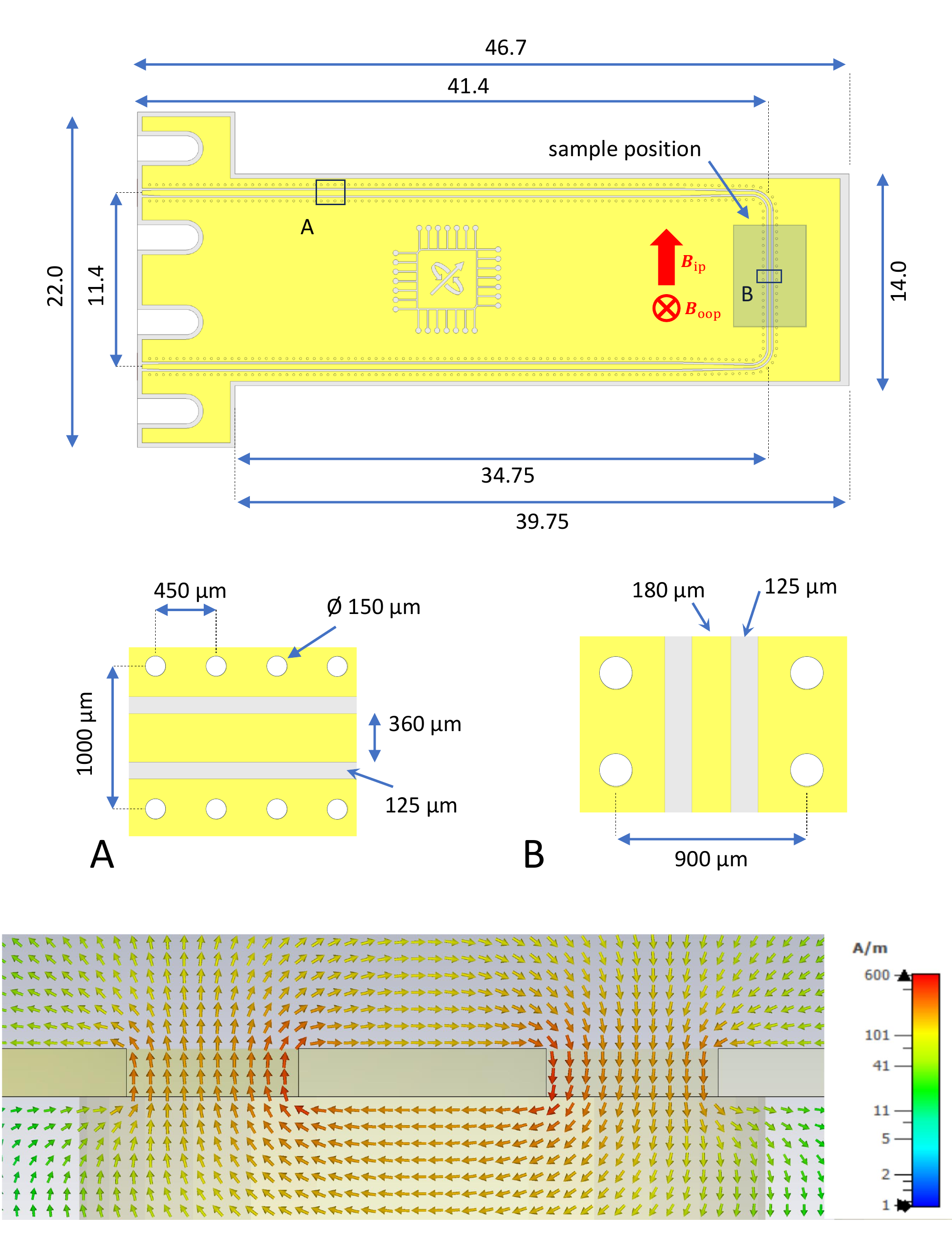}
\caption{\label{fig:GCPW}Technical drawing of the grounded coplanar waveguide. Units are in millimeters if not stated otherwise. Yellow: metal layer, grey: substrate layer, white: vias. Bottom: Maximum magnetic field distribution around the sample position at 22.5\,GHz.}
\end{figure}

\begin{figure}
\includegraphics[width=8.6cm]{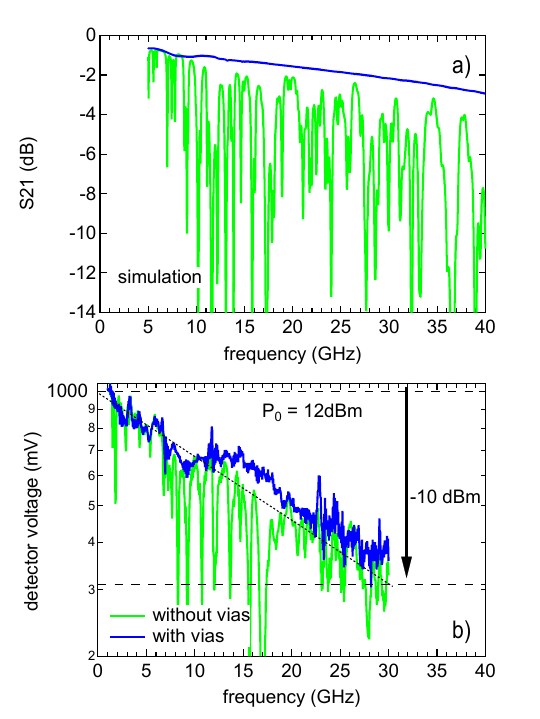}
\caption{\label{fig:frequency_sweeps}a): Simulated frequeny sweeps of the scattering function (transmission) S$_{21}$ for the GCPW without and with via fences. b): Frequency sweeps of the full setup with two different coplanar waveguides, with or without via fences. Notably, the loss at 30\,GHz is determined mostly by the cable losses of approximately 3\,dBm/m. The insertion loss of our GCPW can thus be estimated to approximately 4\,dB. The detector output voltage is consistent with the calibrated input power of approximately 12\,dBm from the signal generator at low frequency, according to the Krytar 203BK datasheet.}
\end{figure}

The coplanar waveguide (CPW) is a critical component of the system. Inspired by various published solutions,\cite{Hu2006, Sain2016, He2016, Gonzalez2018, Psuj2021, Montanheiro2022, Bainsla2022} we designed a simple grounded CPW (GCPW) that is compatible with both the in-plane and out-of-plane measurement geometries and fits any electromagnet with a pole cap diameter of 60\,mm. We started the design from the datasheet of the WithWave 2.92mm edge launch connectors. For in-plane measurements, a waveguide section parallel to the external field is required, so we designed a waveguide with a U-shape and optimized gaps, curve radii, and via fences in CST Studio to minimize impedance mismatches and ensure smooth transitions between different sections of the waveguide. The basic design consists of a GCPW structure created on an FR4-backed 200\,$\mathrm{\mu}$m thick Rogers 4003C substrate with 18\,$\mathrm{\mu}$m Cu layers. The conductor line width is 360\,$\mathrm{\mu}$m, the gap width is 125\,$\mathrm{\mu}$m. This design is adjusted for a 50\,$\Omega$ characteristic impedance. The conductor has a constriction to 180\,$\mathrm{\mu}$m underneath the sample, with smooth transitions between the sections (2\,mm taper length, 1\,mm bend radius). Figure \ref{fig:GCPW} shows the outline and most important lengths of the GCPW. Because of the skin effect, the RF magnetic field is mostly confined to the edges of the inner conductor at high frequencies, but at lower frequencies the constriction provides some magnetic field enhancement, see Fig. \ref{fig:GCPW} (bottom). Furthermore, the narrower conductor limits the spacial extension of eddy currents induced by edge fields, which give rise to asymmetric line profiles.\cite{Flovik2015} Via fences on a GCPW ensure the suppression of transverse resonant modes and thereby keep radiated losses low.\cite{Sain2016} We include blind vias with 150\,$\mathrm{\mu}$m diameter and 450\,$\mathrm{\mu}$m pitch, which ensure high-frequency operation far beyond our design specification of 30\,GHz. They are placed at a distance of 150\,$\mathrm{\mu}$m from the ground-plane edge. According to the $\lambda/4$-criterion, the cutoff frequency can be written as $f_\mathrm{c} = c_0 / (4 d_\mathrm{v} \sqrt{\varepsilon_\mathrm{eff}})$ with the speed of light in vacuum $c_0$, the via pitch $d_\mathrm{v}$, and the effective dielectric constant $\varepsilon_\mathrm{eff} \approx 2.37$ for our GCPW. Thus, we obtain $f_\mathrm{c} \approx 108$\,GHz with our design and ensure its performance if replacement with higher-bandwidth RF components is feasible. Only the connectors will have to be replaced with, e.g., 1.85mm connectors for measurements up to 67\,GHz. Higher frequencies are desirable in FMR measurements to improve the accuracy of the parameter estimates and to enable measurements on films with very large perpendicular magnetic anisotropy. We note that the 2.92\,mm connectors in our basic design are already rated to 40\,GHz. In Figure \ref{fig:frequency_sweeps}, we show the simulated transmission through the waveguide with and without the via fences. Additionally, we show the measured detector voltage of our setup with waveguides with and without via fences. Removing the vias and reducing the waveguide to a conductor-backed CPW (CB-CPW) leads to a sharply fringed transmission function with high losses. Furthermore, erratic behaviour of the line asymmetries, and overall reduced FMR signal magnitude are observed with the CB-CPW and its use is discouraged. The design files are available and can be readily sent to a commercial PCB manufacturer. In the manufacturing process it is mandatory to avoid using a Nickel coating as the adhesive layer before Au deposition. We used an immersion Sn surface coating without a Nickel adhesive layer. Our waveguides were manufactured by Becker \& Müller Schaltungsdruck GmbH in Germany. More details are given in the Supplemental Material.

\begin{figure}
\includegraphics[width=8.6cm]{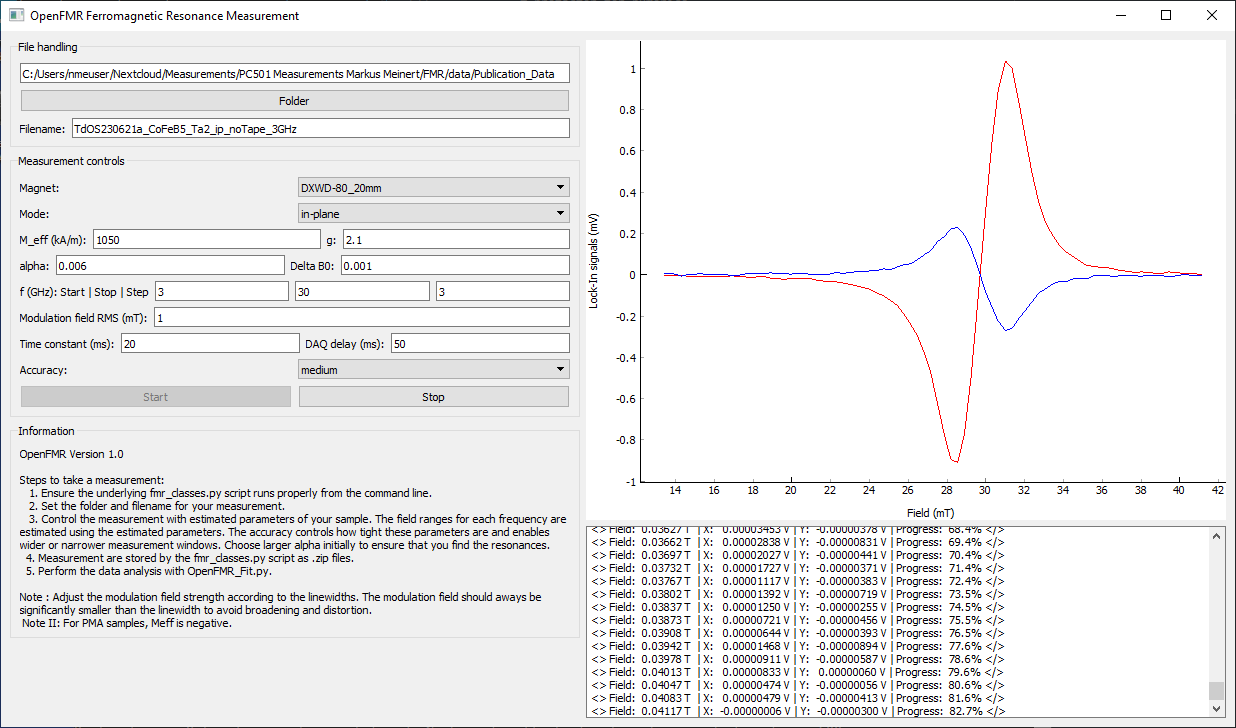}
\caption{\label{fig:DAQ_GUI}Screenshot of the data acqusition GUI. The measurement data presented was obtained on a Co$_{40}$Fe$_{40}$B$_{20}$ 5\,nm thin film deposited on a Si / SiOx wafer.}
\end{figure}

\begin{figure}
\includegraphics[width=8.6cm]{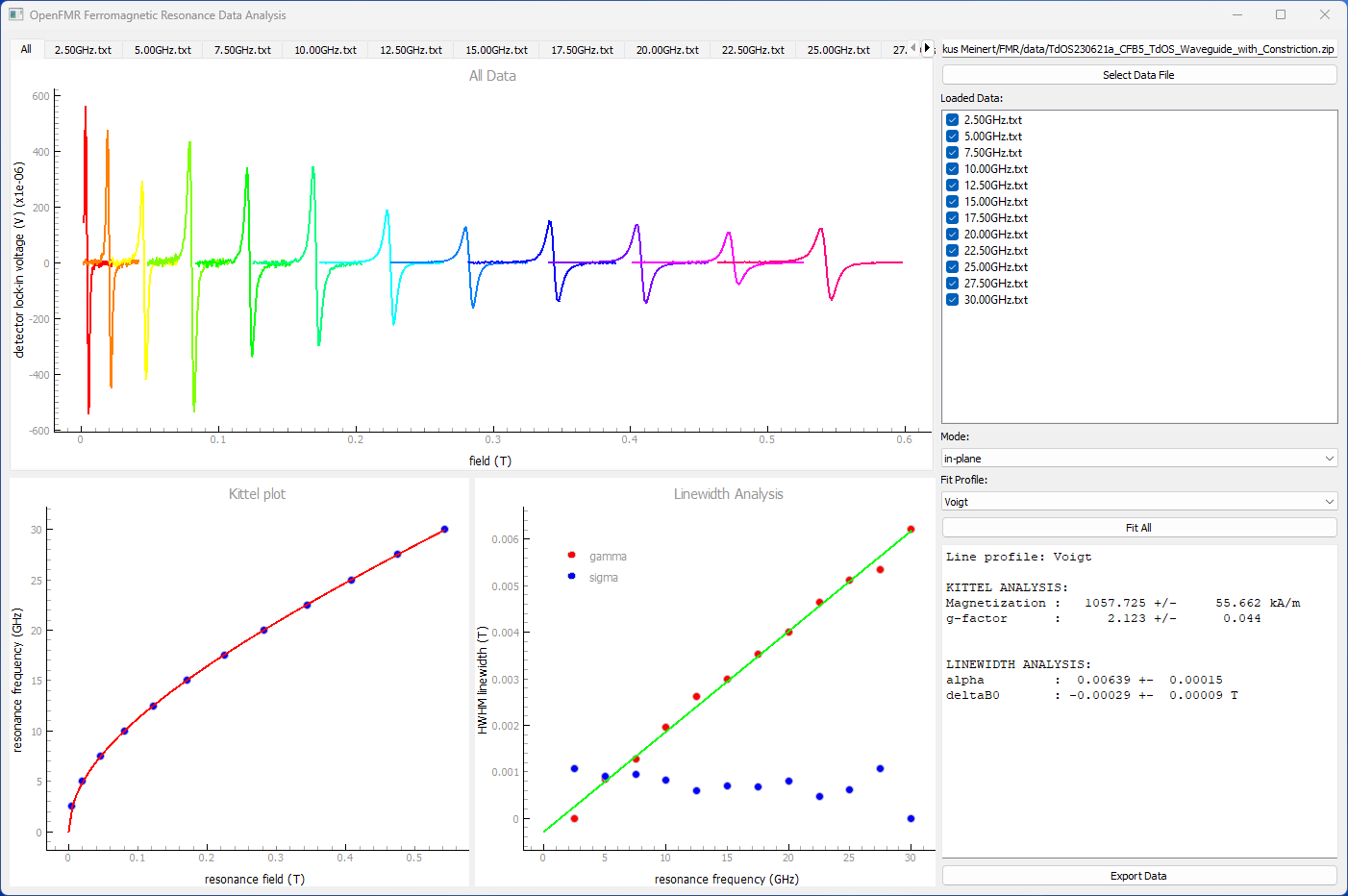}
\caption{\label{fig:data_analysis}Screenshot of the graphical data analysis program. The measurement data presented was obtained on a Co$_{40}$Fe$_{40}$B$_{20}$ 5\,nm thin film deposited on a Si / SiOx wafer.}
\end{figure}

\section{\label{sec:software}Software}
We wrote the software stack in Python 3. The language choice ensures easy code maintenance and extentability, while also being ideally suited for an educational environment. Its main components are
\begin{enumerate}
\item Low-level device interfaces,
\item Magnet control abstraction layer,
\item data acquisition (DAQ) script,
\item graphical user interface for DAQ script,
\item graphical data analysis software.
\end{enumerate} 
The low-level device interfaces control the communication with the various hardware devices via serial and TCP connections and provide a first abstraction layer for higher-level controls. These are imported as a module into the next layer. The magnet control abstraction layer ensures that the system can be controlled in terms of magnetic field values and provides the conversion to power supply currents via interpolated calibrations of the magnet for the two gap values used in our implementation for ip and oop measurements. It further provides a simple interface for the measurement of the field via digital or analog interfaces with the Gaussmeter. The data acqusition script provides a measurement loop, where field-sweeps are performed for multiple frequency values. To facilitate a quick measurement with high resolution, we use parameter estimates provided by the user for the magnetization and the damping together with an accuracy setting, so that the software can perform the sweeps around the estimated resonance fields with high resolution. Compensation of the CPW and cable damping and the lower FMR signal due to Gilbert damping and associated frequency-dependent linewidth increase is done by automatically adjusting the power of the RF signal generator between 0\,dBm at 0\,GHz and 15\,dBm at 30\,GHz. The data is stored as text files within a zip file container.

\begin{figure}
\includegraphics[width=8.6cm]{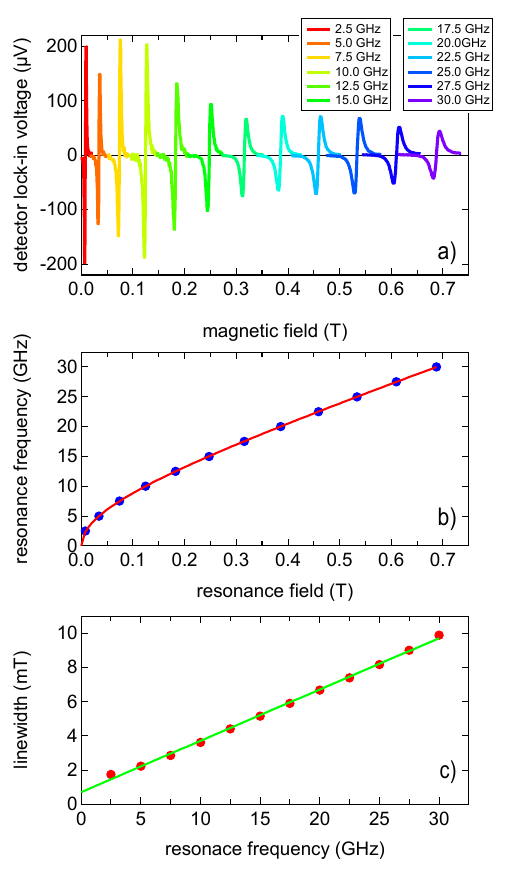}
\caption{\label{fig:NiFe_ip}In-plane FMR measurement of a Si / SiOx / NiFe 5\,nm / Ta 2\,nm thin film sample. The fits were done with the asymmetric Lorentz line profile. The results are $M_\mathrm{eff} = (610 \pm 9)$\,kA\,/\,m, $g=2.143 \pm 0.010$, $\alpha = 0.009 \pm 0.00012$ and $\Delta B(0) = 0.00071$\,T.}
\end{figure}

\begin{figure}
\includegraphics[width=8.6cm]{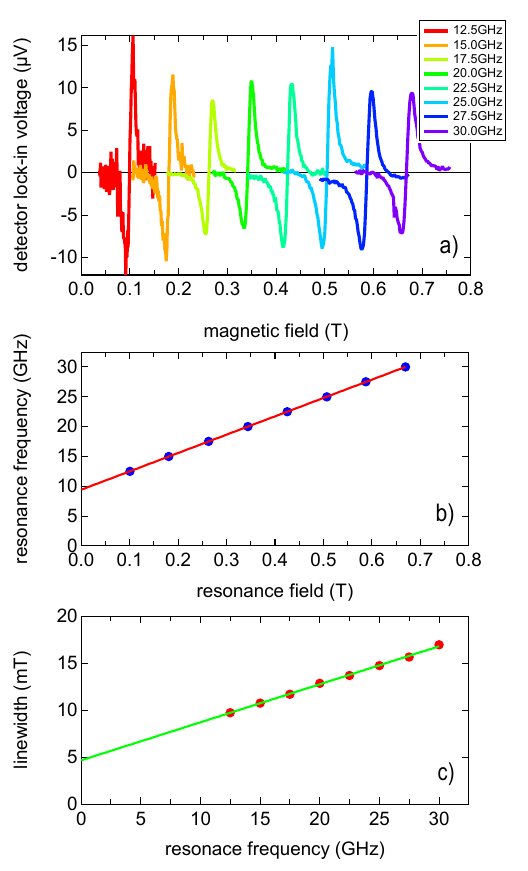}
\caption{\label{fig:CoFeB_oop}Out-of-plane FMR measurement of a Si / SiOx / Ta 8\,nm / CoFeB 1\,nm / MgO 2.5\,nm / TaOx 2.5\,nm (annealed at 200$^\circ$C) thin film sample. The fits were done with the asymmetric Lorentz line profile. The results are $M_\mathrm{eff} = (-243.6 \pm 0.4)$\,kA\,/\,m, $g=2.198 \pm 0.002$, $\alpha = 0.0125 \pm 0.0002$ and $\Delta B(0) = 0.0047$\,T.}
\end{figure}

The graphical user interface (GUI) for the data acqusition provides an easy interface for the measurement script and has a live data preview during the measurement, see the screenshot in Figure \ref{fig:DAQ_GUI}. The graphical data analysis software (see screenshot in Fig. \ref{fig:data_analysis}) reads the measurement files and performs line fits with a choice of line profiles: Lorentz, Asymmetric Lorentz, or Voigt. The Lorentz line profile is given by the field derivative of equation \eqref{eq:Lorentz}. The Asymmetric Lorentz line shape arises from eddy currents in the sample that give rise to an out-of-phase excitation of the magnetization.\cite{Flovik2015, Gladii2019} The line shape is given by a Dyson line profile\cite{Popovych2015}
\begin{equation}
\Delta P_\mathrm{D}(B) = A \left( 1 + \frac{2 \beta (B_\mathrm{res} - B) }{\Gamma_B}\right) \frac{\Gamma_B^2}{\left(B_\mathrm{res} - B\right)^2 + \Gamma_B^2 }.
\end{equation}
The parameter $\beta$ scales the out-of-phase contribution to the line profile. The fits are performed with the numerical derivative of $\Delta P_\mathrm{D}(B)$. Finally, we can use a Voigt profile for the line fit.\cite{Kupriyanova2016} It is defined as the convolution of a Lorentz profile with a Gaussian (with a standard deviation of $\sigma$), and is useful for describing the additional broadening which originates from the modulation field. This will mostly play a role at low frequencies, because the line width is given as\cite{Olivero1977} $\Delta B_\mathrm{FWHM} \approx \Gamma + \sqrt{\Gamma_B^2+ 8 \mathrm{ln}(2) \,\sigma^2}$ and constant $\sigma$. The fits are again performed with the numerical derivative of the profile. Subsequently, fits of the resonance positions are done with the Kittel equations \eqref{eq:Kittel_ip} or \eqref{eq:Kittel_oop} and of the linewidths according to equation \eqref{eq:width}. More complex situations of, e.g. in-plane anisotropy require specific extensions with the corresponding Kittel equations and may require multiple measurements under different directions for a complete analysis. This is out of the scope of the present version of our software. In this case, it is advisable to use the extracted resonance fields and continue the data analysis with a dedicated external script. The same holds true for the implementation of an asymptotic fit parameter analysis.\cite{Shaw2013} Analysis results including the fits can be exported and stored inside the measurement zip files. Individual frequencies can be selected or deselected for the fit to observe the influence on the results individually or to remove outliers. 

The graphical user interfaces for both the DAQ and the data analysis programs were written with the help of ChatGPT-4. This allowed us to focus on the development of the underlying algorithms, while the user interface was generated iteratively using an interactive chat with the large language model. We use the PyQT 5 framework, the pyqtgraph library, and numerous standard modules.

\section{\label{sec:examples}Examples}
Here, we present two examples measured with the presented setup. For the demonstration of the in-plane measurement, we use a film stack of Si / SiOx / Ni$_{80}$Fe$_{20}$ 5nm / Ta 2nm, made with dc magnetron sputter deposition. For the out-of-plane measurement demonstration, we use a film stack of Si / SiOx / Ta 8nm / Co$_{40}$Fe$_{40}$B$_{20}$ 1nm / MgO 2.5nm / TaOx 2.5nm deposited by dc and rf magnetron sputtering. The film was post-annealed at 200$^\circ$C to achieve the perpendicular magnetic anisotropy. The substrates were cleaved into $5 \times 5$\,mm$^2$ pieces. Data were collected with a lock-in time constant of 20\,ms (300\,ms integration time) for the ip measurement and 100\,ms (1500\,ms integration time) for the oop measurement with 4th order low pass filters. The modulation field was 1\,mT and 2\,mT RMS, respectively, with a frequency of 423\,Hz. The data, shown in Figures \ref{fig:NiFe_ip} and \ref{fig:CoFeB_oop}, shows the good quality of the measured raw data (panels a)). Noise seen in the low-frequency measurements of the oop experiment originates from the signal generator, which becomes much less noisy above 15\,GHz. However, this issue is only significant for extremely thin films, such as our just 1\,nm thick CoFeB layer. The corresponding Kittel and linewidth plots are shown along with their respective fits to equations \ref{eq:Kittel_ip}, \ref{eq:Kittel_oop}, and \ref{eq:width} in panels b) and c), respectively. Extracted data for $M_\mathrm{eff}$, $g$, $\alpha$, and $\Delta B_0$ are given in the respective captions. We note that the typical data collection time for a single field-sweep is about 100\,s.

\section{Conclusion}
In summary, we present a relatively cheap broadband FMR spectrometer for educational or general materials characterization applications. It is capable of detecting the ferromagnetic resonance in ultrathin magnetic films with perpendicular magnetic anisotropy. We provide detailed descriptions for how to build the system, along with the designs for the CPW, the holders and support structures for 3D printing, and the complete Python code. All designs and software are published under the GNU GPL v3.0 license.

\begin{acknowledgments}
This work was partially supported by the Deutsche Forschungsgemeinschaft under Project Numbers 468939474, 511340083 and 513154775. 

\end{acknowledgments}

\section*{AUTHOR DECLARATIONS }

\subsection*{Conflict of Interest}
The authors have no conflicts to disclose.

\subsection*{Author Contributions}
\textbf{Markus Meinert}: Conceptualization; Data curation (equal); Formal analysis (lead); Funding acquisition; Investigation (lead); Methodology (lead); Project administration; Software (lead); Supervision (lead); Validation (equal); Visualization (equal); Writing – original draft; Writing – review \& editing (equal).

\textbf{Tiago de Oliveira Schneider}: Data curation (equal); Formal analysis (supporting); Investigation (supporting); Methodology (supporting); Resources (lead); Software (supporting); Supervision (supporting);Validation (equal); Visualization (equal); Writing – review \& editing (equal).

\textbf{Shalini Sharma}: Resources (supporting); Writing – review \& editing (equal).

\textbf{Amir Khan}: Resources (supporting); Writing – review \& editing (equal).

\section*{Data Availability Statement}

The data that support the findings of this study are available upon reasonable request from the corresponding author. The software and design files are available at https://github.com/MeinertTUDa/OpenFMR.

\section*{References}

\bibliography{cite} 

\end{document}